\documentstyle[amssymb,twocolumn,prl,aps,epsf]{revtex}

\begin{document}          

\twocolumn[\hsize\textwidth\columnwidth\hsize\csname
@twocolumnfalse\endcsname

\title{Many-body theory of the quantum mirage}
\author{A. A. Aligia}
\address{Comisi\'on Nacional de Energ{\'\i}a At\'omica,\\
Centro At\'omico Bariloche and Instituto Balseiro,\\
8400 S.C. de Bariloche, Argentina.}
\date{Received \today }
\maketitle

\begin{abstract}
In recent scanning tunneling microscopy experiments, confinement in an
elliptical corral has been used to project the Kondo effect from one focus
to the other one. I solve the Anderson model at arbitrary temperatures, for
an impurity hybridized with eigenstates of an elliptical corral, each of
which has a resonant level width $\delta $. This width is crucial. If $%
\delta <20$ meV, the Kondo peak disappears, while if $\delta >80$ meV, the
mirage disappears. For particular conditions, a stronger mirage with the
impurity out of the foci is predicted.
\end{abstract}

\pacs{Pacs Numbers: 72.15.Qm, 68.37.Ef, 73.20.Fz}

]

\narrowtext

The recent advances in nanofabrication has led to fascinating experiments.
In one of them, the electrons on the Cu(111) surface  are confined in a
quantum corral assembled by depositing other atoms on the boundary of an
ellipse. When a Co atom is positioned at one focus of the ellipse, the
corresponding Kondo signature is observed by scanning tunneling microscopy
(STM) not only at the atom, but also at the other focus.\cite{mano,fie}
Another remarkable feature of the experiments is that eigenstates of a
two-dimensional free electron gas confined into the ellipse are clearly
displayed. The average separation between the corresponding eigenvalues is $%
\sim 10$ meV, and one might expect that the confinement is good enough such
that the level width of each eigenstate $\delta \lesssim 10$ meV. On the
other hand, a Kondo peak in the impurity spectral density of states is in
general observed when the impurity hybridizes with a continuum of conduction
states.\cite{cos} Can the Kondo feature exist if these states are discrete?
From recent work in mesoscopic systems, we know that the answer is positive
if the Kondo temperature $T_{K}\gtrsim d$, where $d$ is the average
separation of the levels which have a significant hybridization with the
impurity.\cite{thi} However, in the experiments $T_{K}\sim 5$ meV, while $%
d\sim 100$ meV.

In this Letter, I calculate the impurity and conduction density of states as
a function of $\delta $. This requires an explicit calculation of the
many-body effects, which has not been done in previous theories of the
mirage effect.\cite{fie,aga,por,wei} I find that only for values of $\delta $
large enough to lead to a well defined Kondo peak, but low enough to allow
coherence between both foci, the theory is consistent with experiment. From
the insight gained on the decoherence effects, I propose a slight
modification of the experiment, which should result in an enhanced mirage
effect, with the impurity out of the foci.

To describe the essential features of the experiment, I use the following
Anderson model:\cite{por} 
\begin{eqnarray}
H &=&\sum_{j\sigma }\varepsilon _{j}c_{j\sigma }^{\dagger }c_{j\sigma
}+E_{d}\sum_{\sigma }d_{\sigma }^{\dagger }d_{\sigma }+Ud_{\uparrow
}^{\dagger }d_{\uparrow }d_{\downarrow }^{\dagger }d_{\downarrow }  \nonumber
\\
&&+\sum_{j\sigma }V_{j}(c_{j\sigma }^{\dagger }d_{\sigma }+\text{H.c.}%
)+H^{\prime }.  \label{ham}
\end{eqnarray}
Here $c_{j\sigma }^{\dagger }$ and $d_{\sigma }^{\dagger }$ create an
electron on the $j^{th}$ eigenstate of a hard wall elliptic corral and the
impurity respectively. Roughness of the boundaries of the corral on a length
scale smaller than the Fermi wave length $2\pi /k_{F}\sim $ 30\AA\ does not
affect much the wave functions.\cite{mano,fie} $H^{\prime }$ describes the
hopping of each corral state with a continuum of conduction states outside
the corral, with the same symmetry. If the hopping and spectral density of
these states is constant, and if the effective interaction between corral
states can be neglected, the effect of $H^{\prime }$ is simply to introduce
a width $\delta $ in the unperturbed ($V_{j}=0$) Matsubara Green function of
each state:

\begin{equation}
G_{j}^{0}(i\omega _{n})=\langle \langle c_{j\sigma };c_{j\sigma }^{\dagger
}\rangle \rangle _{i\omega _{n}}=[i\omega _{n}-\varepsilon _{j}+i\delta 
\text{sgn}(\omega _{n})]^{-1}.  \label{gc0}
\end{equation}
I take this form with $\delta $ independent of $j$ for simplicity. $V_{j}$
should be proportional to the (real) $j^{th}$ normalized wave function $%
\varphi _{j}(r)$ of the corral at the impurity position $R_{i}$. I take:

\begin{equation}
V_{j}=25\text{ meV }\sqrt{ab}\varphi _{j}(R_{i})\max \left( 1+\frac{%
\varepsilon _{F}-\varepsilon _{j}}{\text{eV}},0\right) .  \label{v}
\end{equation}
$a$ ($b$) is the semimajor (semiminor) axis of the ellipse, and $\varepsilon
_{F}$ is the Fermi energy. The last factor produces a slow decrease of $V_{j}
$ with energy and a smooth cutoff 1 eV above $\varepsilon _{F}$. It leads to
a more symmetrical line shape, equilibrating approximately the weight of the 
$V_{j}$ at both sides of $\varepsilon _{F}$ and does not affect the physics.
The energy prefactor was chosen to lead to the experimental width of the
Kondo feature for $\delta =50$ meV. If the Cu surface is represented by a
tight binding model with one orbital per site,\cite{wei} this amounts to a
hopping $t^{\prime }\sim 0.67$ eV between the impurity and the atom below it
(or $t^{\prime }/n$ if the impurity has $n$ nearest neighbors on the
surface).\cite{note} This value is reasonable for a hopping between 4s and
3d electrons. The rapid decay of this hopping with distance allows to
neglect hopping of the impurity with other atoms. Since in addition the
density of s and p states at $\varepsilon _{F}$ is two times larger in the
clean surface  than in the bulk,\cite{euc} one can safely neglect the
coupling of the latter with the impurity.

The eigenstates of free electrons inside an ellipse were obtained solving
the matrix Hamiltonian in the basis of eigenstates for a circle, deformed to
fit into the ellipse.\cite{nak} These basis states are proportional to $%
J_{k}(\gamma _{kn}\rho )(e^{ik\theta }\pm e^{-ik\theta })$, where $\rho
=[(x/a)^{2}+(y/b)^{2}]^{1/2}$, $\theta =\arctan [ya/(xb)]$, $J_{k}$ is the $%
k^{th}$ Bessel function and $\gamma _{kn}$ is its $n^{th}$ zero. We retained
the 100 lowest states in each of the four symmetry sectors of the group $%
C_{2v}$. From the solution with an enlarged basis with a total of 1600
states, the error in the $\varepsilon _{j}$ near the Fermi energy is
estimated $\sim 10^{-6}$ eV. Recently, this problem was solved analytically.%
\cite{por} We take an ellipse with eccentricity $e=1/2$, $a=71.3$ \AA\ ($b=%
\sqrt{3}a/2$), and the effective mass was adjusted so that $\varepsilon
_{42}=\varepsilon _{F}$.

The particular structure of the non-interacting problem introduces technical
difficulties in the many-body problem. In addition, the fact that the
temperature of the experiment $T=4$ K $\ll T_{K}=53$ K, seems to invalidate
the use of the non-crossing approximation, since it works for $T\gtrsim
T_{K} $ and violates Fermi liquid relations at low $T$.\cite{thi,schi} I
obtain the retarded Green function for the impurity $G_{d}(\omega )=\langle
\langle d_{\sigma };d_{\sigma }^{\dagger }\rangle \rangle _{\omega }$ using
finite-$T $ perturbation theory up to second order in $U$ and analytic
continuation:\cite{hor} $G_{d}^{-1}=(G_{d}^{0})^{-1}-\Sigma $, with $%
G_{d}^{0}(i\omega _{n})=[i\omega _{n}-\tilde{E}_{d}-%
\sum_{j}V_{j}^{2}G_{j}^{0}(i\omega _{n})]^{-1}$, $\tilde{E}%
_{d}=E_{d}+U\langle d_{\sigma }^{\dagger }d_{\sigma }\rangle $, and:

\begin{equation}
\Sigma (i\omega _{l})=-(UT)^{2}\sum_{n,m}G_{d}^{0}(i\omega _{l}-i\nu
_{m})G_{d}^{0}(i\omega _{n})G_{d}^{0}(i\omega _{n}+i\nu _{m}),  \label{sigma}
\end{equation}
with $\omega _{n}=(2n+1)\pi T$ and $\nu _{m}=2m\pi T$. For cases studied
before (with a simple band structure), comparison with calculations using
Wilson's renormalization group\cite{cos} shows that the approximation is
qualitatively correct for all values of $U$. The method has proven to be
quantitatively correct even for $U=2.5\pi \Delta $, where $\Delta $ is the
resonant level width.\cite{hor} In the present case, the particular features
of the non-interacting system renders an estimate of $\Delta $ difficult,
but in absence of the corral $\Delta \sim 0.2$ eV.\cite{ujs} I have chosen $%
U=1$ eV. This value is enough to lead to a strong Kondo peak, remaining
within the range of validity of the approximation. $E_{d}$ was adjusted in
order that the Hartree-Fock effective d level $\tilde{E}_{d}$ falls near $%
\varepsilon _{F}$, as suggested by first-principles calculations,\cite
{wei,wei2} with a small shift of -22 meV to displace the Kondo feature to
the experimental position for $\delta =50$ meV. The resulting impurity
spectral function $\rho _{d}(\omega )=-%
\mathop{\rm Im}%
\{G_{d}(\omega )\}/\pi $ is represented in Fig. 1 (a) for $\delta =50$ meV.
The shifts in the broad peaks from $E_{d}\sim -0.5$ eV and $E_{d}+U\sim 0.5$
eV are due to the particular structure of $\varepsilon _{j}$ and $V_{j}$.
The only peak of interest here is the central one. Its temperature
dependence is shown in Fig. 1 (b). The peak looses intensity and broadens as
the temperature is increased. This is a clear signature of its many-body
nature. For $T=T_{K}$, the maximum of the peak with respect to the
background is reduced to roughly half its value at $T=0$.

Once $G_{d}(\omega )$ is known, the local conduction electron retarded Green
function $G_{c}(r,\omega )=\langle \langle \psi (r);\psi (r)^{\dagger
}\rangle \rangle _{\omega }$, with $\psi (r)=\sum_{j}\varphi _{j}(r)c_{j}$,
can be obtained using the following exact relationship (simplified for real $%
\varphi _{j}(r)$) obtained from the equations of motion:

\begin{equation}
G_{c}(r,\omega )=\sum_{j}\varphi _{j}^{2}(r)G_{j}^{0}(\omega )+G_{d}(\omega
)\{S(r,\omega )\}^{2},  \label{gc}
\end{equation}
with

\begin{equation}
S(r,\omega )=\sum_{j}\varphi _{j}(r)V_{j}G_{j}^{0}(\omega ).  \label{s}
\end{equation} 

\begin{figure}
\narrowtext
\epsfxsize=3.5truein
\vbox{\hskip 0.05truein \epsffile{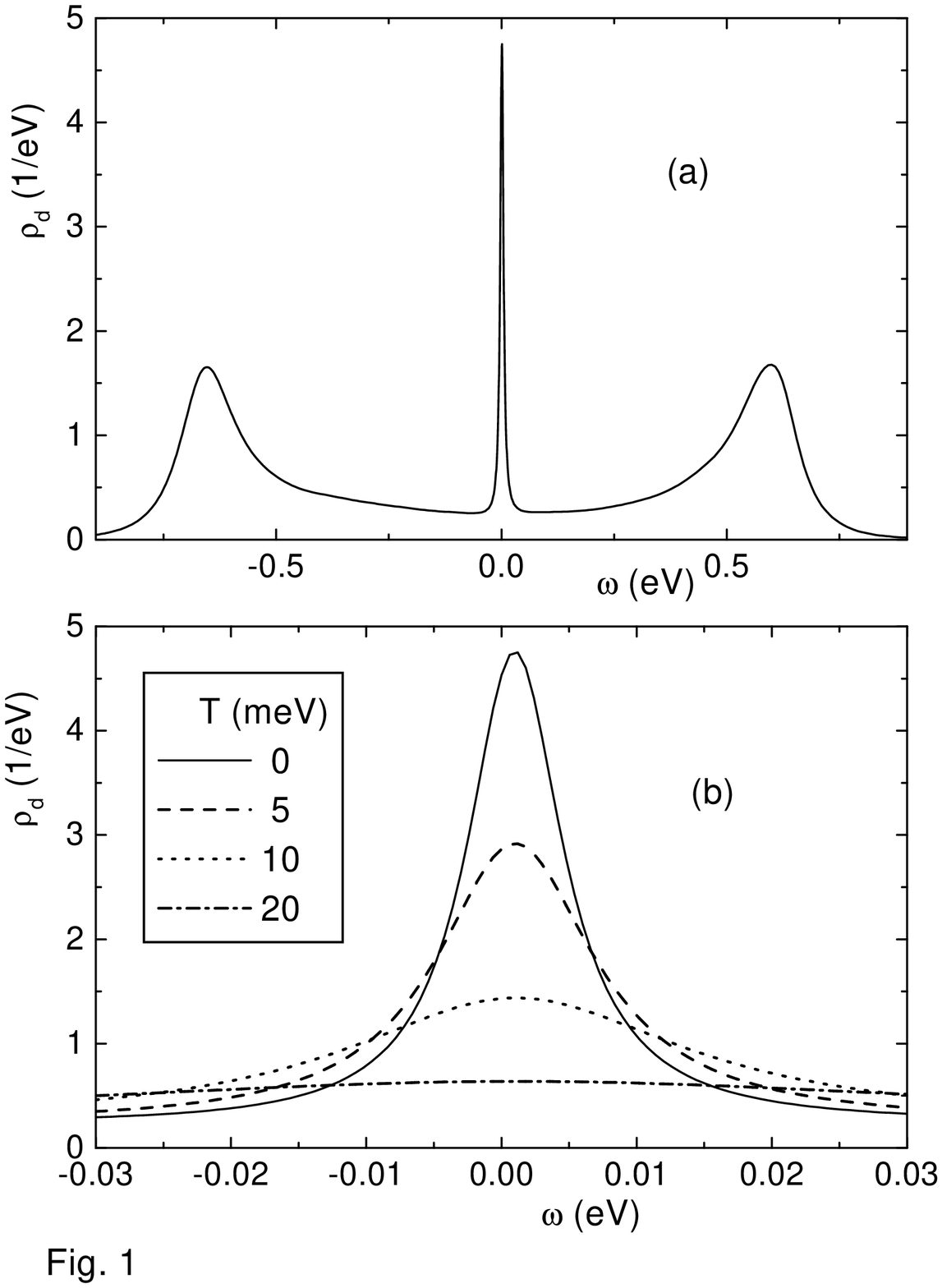}}
\medskip
\caption{(a) Impurity spectral density as a function of
energy. (b) Temperature dependence of the central peak.}
\end{figure}

The differential tunneling conductance $dI/dV$ is essentially proportional
to the local conduction spectral density $\rho _{c}(r,\omega )=-%
\mathop{\rm Im}%
\{G_{c}(r,\omega )\}/\pi $. At or very near $R_{i}$ there might be a small
deviation due to direct tunneling between the STM tip and the impurity,
which affects the line shape.\cite{schi,ujs} This direct contribution cannot
be transmitted to the mirage and for the sake of clarity, I neglect it here.
In addition, the similarity between the observed line shape at both foci
suggests that this effect can be neglected.

For the case of $\varphi _{j}(r)V_{j}$ independent of $j$ and constant
density of $\varepsilon _{j}$ (as in the case of realistic models for 4f
impurities in a metallic bulk\cite{cos} with $r$ at the impurity site), the
sum $S(r,\omega )$ can be evaluated easily, and in usual cases in which the
distribution of $\varepsilon _{j}$ is much broader than the peak in $\rho
_{d}(\omega )$, the resulting $S$ lies near the imaginary axis for $\omega $
near the peak, and $S^{2}$ can be approximated by a negative real constant.
Then, the change in $\rho _{c}(r,\omega )$ after addition of the impurity $%
\Delta \rho _{c}=-%
\mathop{\rm Im}%
\{G_{d}S^{2}\}/\pi $, is proportional to -$\rho _{d}(\omega )$ near its
peak. Physically, the level repulsion with the impurity states causes a
depression of $\rho _{c}(r,\omega )$. However, the above mentioned
hypothesis are not valid in the present case, and although this picture
remains qualitatively correct for large $\delta $, the situation is
different for small $\delta $. 

\begin{figure}
\narrowtext
\epsfxsize=3.5truein
\vbox{\hskip 0.05truein \epsffile{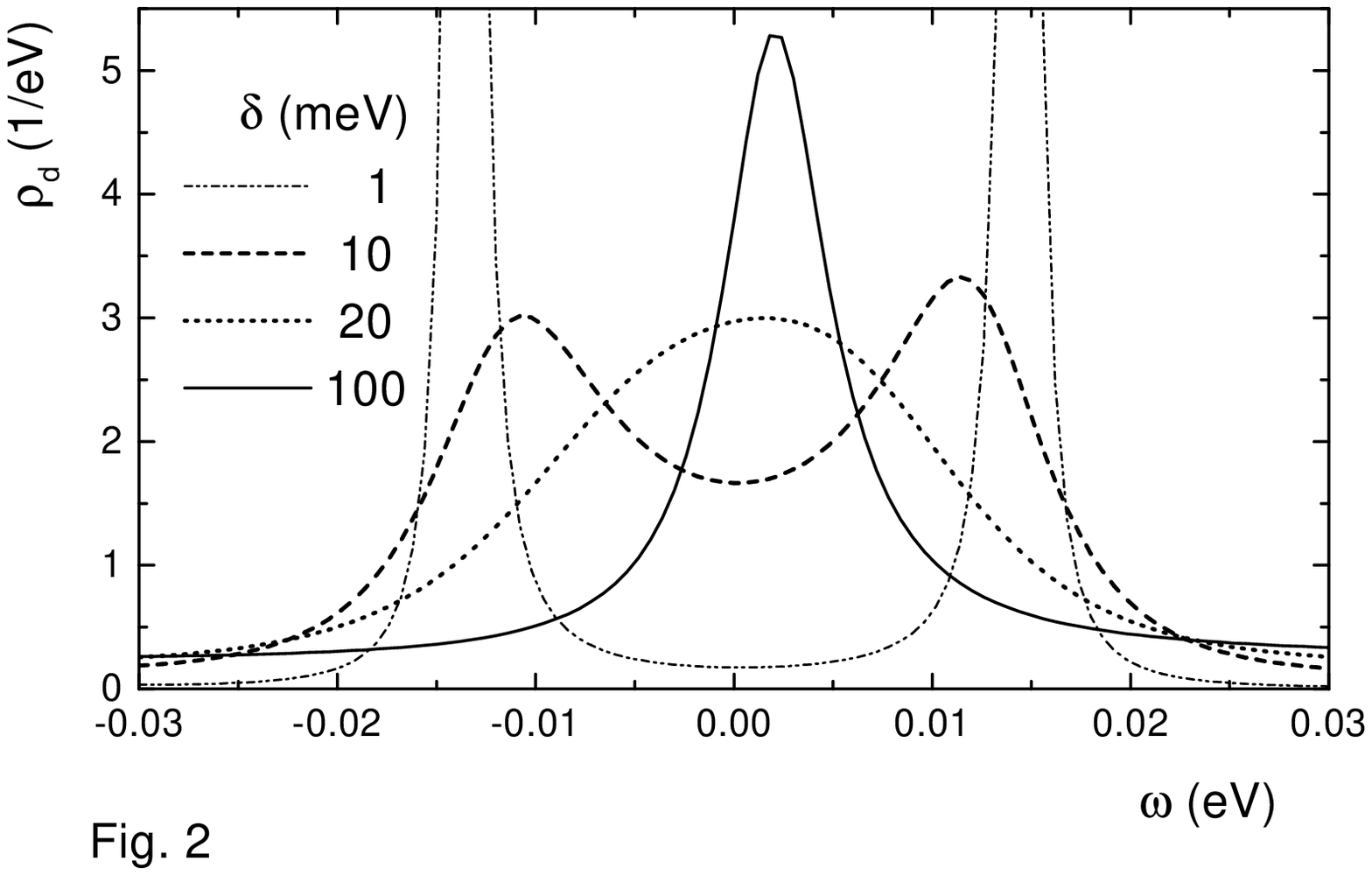}}
\medskip
\caption{Impurity spectral density as a function of energy
for several values of $\delta $.}
\end{figure}

In Fig. 2 we show the dependence of $\rho
_{d}(\omega )$ with $\delta $ at $T=0$. If $\delta $ is very small, there
are two narrow peaks, which correspond to two discrete energy levels for $%
\delta \rightarrow 0$. This agrees with the results of Thimm {\it et al. }%
for the case when one of the discrete states coincides with $\varepsilon
_{F} $ (the figure remains practically unchanged if $\varepsilon _{F}$ is
increased to $\varepsilon _{43}$).\cite{thi} In addition, for $\delta =1$
meV, $\rho _{c}(r,\omega )$ {\em increases }at the position of these two
peaks after adding the impurity, and the shape of $\Delta \rho _{c}(\omega )$%
, although smoother than that of $\rho _{c}(\omega )$ is also in strong
disagreement with experiment. The structure with two peaks in $\rho
_{d}(\omega )$ disappears for $\delta \sim 18$ meV, but a clear sharp peak
does not develop until $\delta \sim 30$ meV. In spite of this, as shown in
Fig. 3 (a), $\Delta \rho _{c}(R_{i},\omega )$ for $\delta =20$ meV is
roughly consistent with experiment. This is easily understood noting that
for small $\delta $, $S$ is dominated by the term proportional to $%
G_{42}^{0}\cong (\omega -\varepsilon _{42})^{-1}$, and $%
\mathop{\rm Im}%
\{G_{d}S^{2}\}$ becomes narrower than $%
\mathop{\rm Im}%
G_{d}.$

The fact that $S$ is dominated by the term with $j=42$, means that
essentially only this state ``feels'' the presence of the impurity at $R_{i}$
and leads to a large mirage effect, since this state transmits coherently
the information to the other focus (see Fig. 3 (a)). However, in the
experiment the intensity at the empty focus is nearly 8 times smaller. As $%
\delta $ increases, other conduction states increase their contribution to $%
S $ and $\rho _{c}$. By symmetry, all of them are even under reflection
through the major axis, but they can have any parity under reflection
through the minor axis $\sigma $. While $\varphi _{42}(r)$ is even under $%
\sigma $, the other states nearest to $\varepsilon _{F}$ with important
hybridization with the impurity ($|V_{j}|>|V_{42}|/2$), namely 32, 35 and
51, are all odd. These states contribute with the same sign as 42 to the
imaginary part of $S$ at the impurity focus $R_{i}$ (see Eqs. (\ref{v}) and (%
\ref{s})), but with opposite sign at the empty focus $\sigma R_{i}=-R_{i}$.
In other words, at the position of the mirage there is a negative
interference of the states with opposite parity which tends to destroy
coherence. For $\delta =0.1$ eV, the mirage is lost.

\begin{figure}
\narrowtext
\epsfxsize=3.5truein
\vbox{\hskip 0.05truein \epsffile{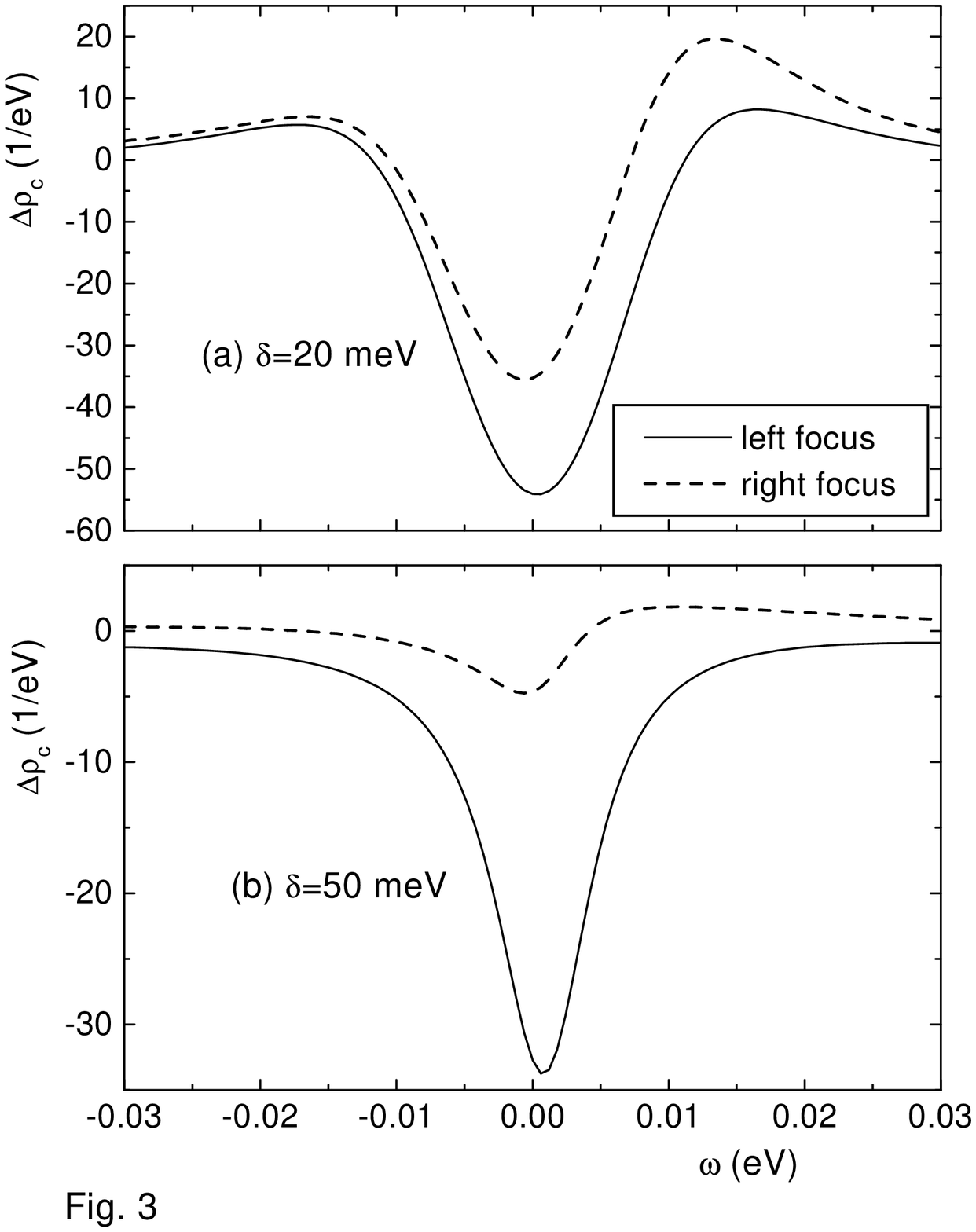}}
\medskip
\caption{Change in the local conduction density of states at
the impurity site (full line) and at the other focus (dashed line) for two
values of $\delta $.}
\end{figure}

To confirm this physical picture, I have repeated the calculation for $%
\delta =0.1$ eV, setting $V_{j}=0$ for $j=$32, 35 and 51. The Kondo feature
becomes significantly narrower and the mirage is restored with an intensity
ratio 1/2. From results similar to those shown in Figs. 3 and 4 as a
function of $\delta $, the experimental situation seems to correspond to $%
\delta \sim 40$ meV. If the direct tunneling between tip and impurity is
indeed important, $\delta $ might be smaller. I have verified that $S$
approaches an imaginary constant for $\delta \geq 50$ meV, and that (as a
consequence) $\Delta \rho _{c}(\omega )$ is quite similar to $\rho
_{d}(\omega )$ inverted for $\delta =50$ meV and all temperatures displayed
in Fig. 1 (b).

In Fig. 4, I show the spatial dependence of $\rho _{c}$ and $\Delta \rho
_{c} $. The former displays some features of the state 43 and is similar to
the experimental topograph (Fig. 2 (c) of Ref. \cite{mano}). However, it is
more sensitive to $\omega $ and $\varepsilon _{F}$ than $\Delta \rho _{c}$,
and to compare with experiment it is necessary to perform an integral over $%
\omega $.\cite{fie} In the difference plot $\Delta \rho _{c}(r)$, the
contribution of states which do not hybridize with the impurity (like 43) is
removed, and as in the experiment, $\Delta \rho _{c}(r)$ displays
essentially minus the density of the state 42. One effect of increasing $%
\delta $ is that the minimum of $\Delta \rho _{c}(r)$, which (as the maximum
of $|\varphi _{42}(r)|^{2}$) is slightly out of focus, displaces towards it
due to the increasing influence of other states.

\begin{figure}
\narrowtext
\epsfxsize=3.5truein
\vbox{\hskip 0.05truein \epsffile{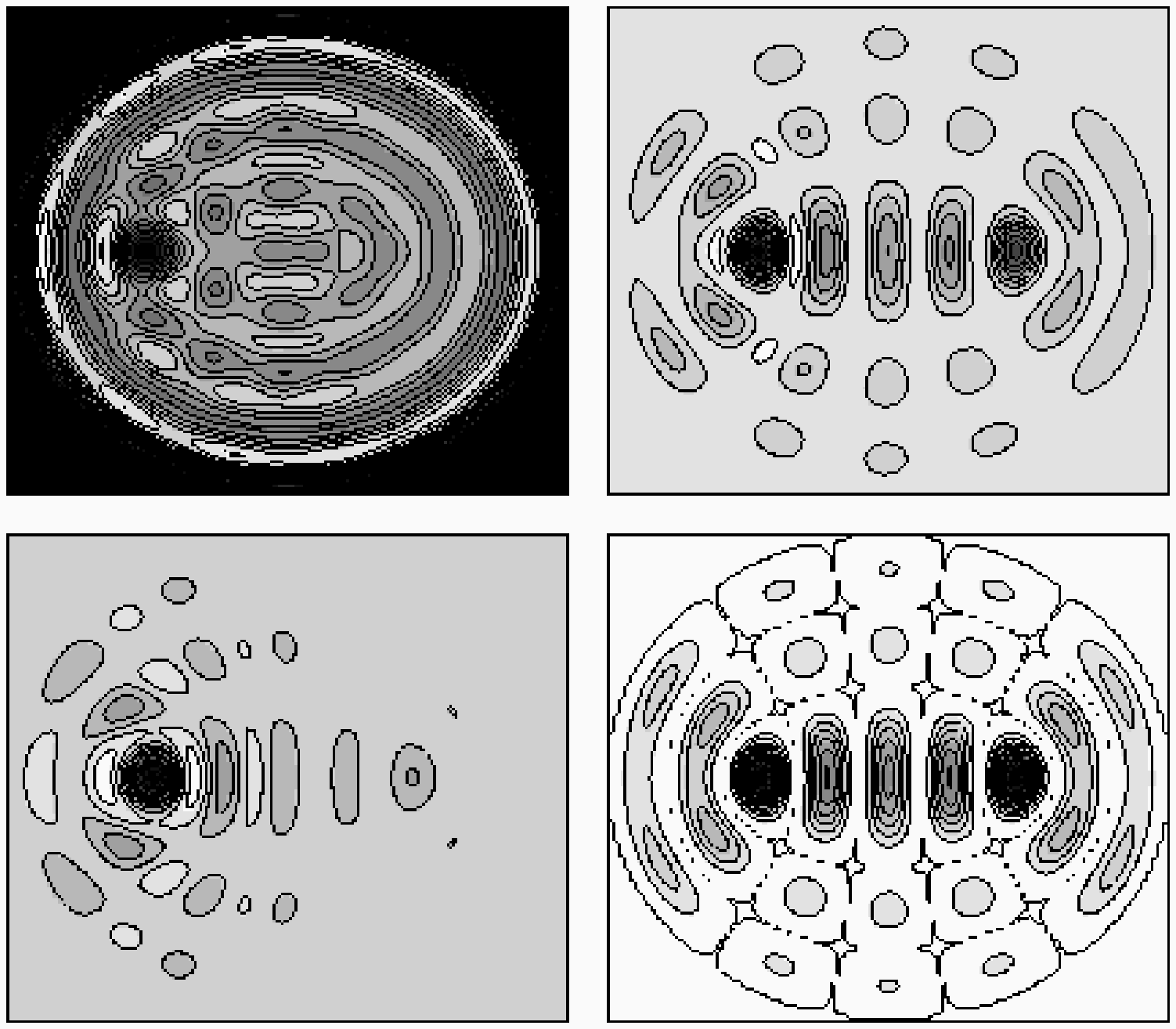}}
\medskip
\caption{Contour plot of $\rho _{c}(r,\omega )$ for $\delta
=50$ meV (top left), $\Delta \rho _{c}(r,\omega )$ for $\delta =20$ meV (top
right) and $\Delta \rho _{c}$ for $\delta =50$ meV (bottom left) , with $%
\omega =10$ meV. For comparison, $-|\varphi _{42}(r)|^{2}$ is displayed at
the bottom right.}
\end{figure}

The analysis of the decoherence effects point out the conditions to improve
the signal at the mirage, in addition to control of the confinement. Keeping
for simplicity $R_{i}$ on the major axis and the mirage point at $\sigma
R_{i}=-R_{i}$, one should find a wave function $\varphi _{l}(r)$ with large
amplitude at some point $R_{i}$, with the condition that all wave functions $%
\varphi _{j}(r)$ with opposite parity under $\sigma $ than $\varphi _{l}(r)$
and $|\varepsilon _{j}-\varepsilon _{l}|\lesssim 0.1$ eV have small
amplitude at $\pm R_{i}.$ The size of the ellipse is then changed keeping $e$
constant to set $\varepsilon _{l}$ near $\varepsilon _{F}$ and the impurity
is positioned at $R_{i}$. It turned out to be difficult to find an ideal
case for conditions close to the experimental ones. However, inspection of
several tenths of wave functions suggests that if $a$ and $b$ are reduced by
a factor 1.104 (so that $a$ becomes 64.6 \AA ) in order that state 35
reaches the Fermi energy, and the impurity is placed at a distance of $0.4a$
from the center of the ellipse, the feature observed at $\sigma R_{i}$
should be larger than that of the original experiment. Keeping other
parameters as in Fig. 3 (b), the resulting $\Delta \rho _{c}(\pm
R_{i},\omega )$ is shown in Fig. 5. The signal at the mirage is in fact
nearly two times larger than in the previous case. The partial loss of
coherence is here due to the states 28 and 42, which are even under $\sigma $%
, while 35 is odd. This result seems hard to explain in terms of billiard
scattering.

\begin{figure}
\narrowtext
\epsfxsize=3.5truein
\vbox{\hskip 0.05truein \epsffile{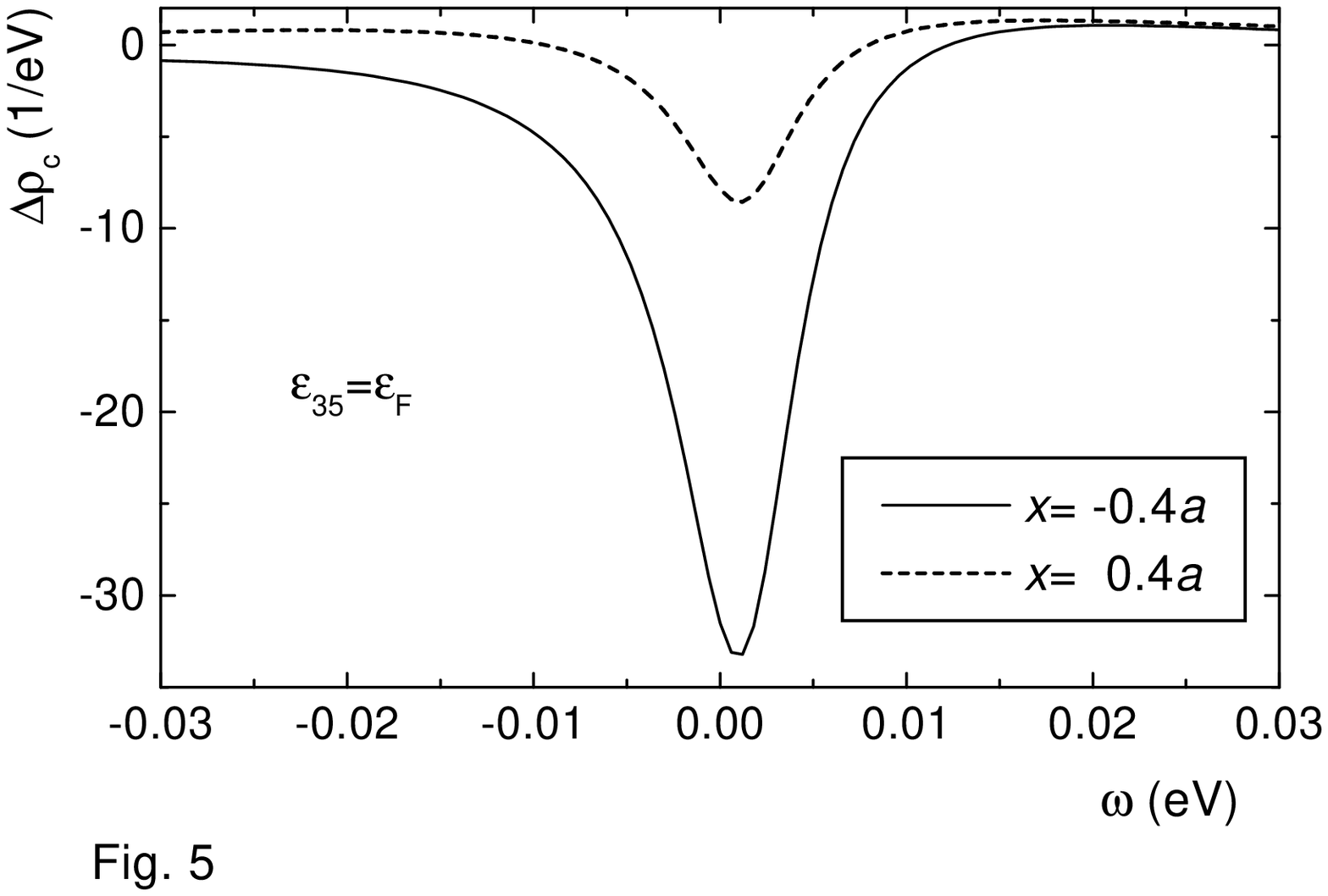}}
\medskip
\caption{$\Delta \rho _{c}(r,\omega )$ {as a function of }$%
\omega $ at two positions $r=(x,0)$, for an ellipse with $\varepsilon
_{35}=\varepsilon _{F}$, $\delta =50$ meV, and the impurity positioned at $%
R_{i}=(-0.4a,0)$.}
\end{figure}

In order to investigate if non-magnetic impurities like S or Si might lead
to a mirage effect, I have repeated some calculations for $U=0$ and tuning $%
E_{d}$ so that $\rho _{d}(\omega )$ is peaked near $\varepsilon _{F}$. In
agreement with Ref. \cite{wei}, a mirage effect roughly consistent with
experiment is possible only if the $V_{j}$ are reduced by a factor $\sim 5$,
so that the hopping $t^{\prime }\sim 0.1$ eV. This seems unrealistic since
replacement of 3d by s or p orbitals is expected to increase the hopping and
involve more sites on the surface.\cite{note} In addition, the property of
the Kondo resonance to tune itself near $\varepsilon _{F}$ is lost.

\noindent

In summary, I have shown that the level width of the conduction states in a
quantum corral, determined by the quality of the confinement, controls the
shape of the observed Kondo feature and the intensity of the mirage effect.
The nature of the decoherence effect is clarified and the conditions to
observe an enhanced effect with the impurity out of the foci are stated.
Replacement of the Kondo resonance by a phenomenological form or a
low-energy phase shift might be valid for poor confinement, but for the
general description, a genuine many-body treatment is needed. To my
knowledge, this is the first calculation of this type for the quantum mirage.

I am indebted to C.R. Proetto for explaining to me the procedure used to
find the eigenstates of free electrons inside an ellipse.\cite{nak} Helpful
discussions with C.R. Proetto, M. Weissmann and H. Bonadeo are gratefully
acknowledged. This work benefitted from PICT 03-00121-02153 of ANPCyT
(Argentina), and PIP 4952/96 of CONICET. I am partially supported by CONICET.


\begin{references}
\bibitem{note}  This argument assumes that the surface states are mainly
concentrated at the surface atomic layer.\cite{euc} If this is not the case, 
$t^{\prime }$ can be larger.
\end{references}
\end{document}